\documentclass[aps,prb,showpacs]{revtex4}
\usepackage{psfrag,pstricks,pst-plot}
\usepackage{amsfonts,epsfig}
\begin{document}
\title{Microwave photoconductivity of two-dimensional electron systems with unidirectional periodic 
modulation}

\author{J\"urgen Dietel}
\affiliation{Institut f\"ur Theoretische Physik, Freie Universit\"at Berlin, Arnimallee
14, 14195 Berlin, Germany}
\author{Leonid I.\ Glazman}
\affiliation{W.I.\ Fine Theoretical Physics Institute, University of Minnesota, Minneapolis, Minnesota 55454, USA}
\author{Frank W.J.\ Hekking}
\affiliation{Laboratoire de Physique et Mod\'elisation des Milieux Condens\'es, CNRS \\ Universit\'e 
Joseph Fourier, BP 166, 38042 Grenoble CEDEX 9, France}
\author{Felix von Oppen}
\affiliation{Department of Condensed Matter Physics,
    Weizmann Institute of Science, Rehovot 76100, Israel\\
    Institut f\"ur Theoretische Physik, Freie Universit\"at Berlin, Arnimallee
14, 14195 Berlin, Germany\footnote{Permanent address}}
\date{\today}
\begin{abstract}
 Motivated by the recently discovered microwave-induced
  ``zero-resistance'' states in two-dimensional electron systems, we
  study the microwave photoconductivity of a two-dimensional electron
  gas (2DEG) subject to a unidirectional static periodic potential.
  The combination of this potential, the classically strong magnetic field,
  and the microwave radiation may result in an anisotropic negative
  conductivity of the 2DEG. Similar to the case of a smooth random
  potential, two mechanisms contribute to the negative
  photoconductivity.  The {\it displacement mechanism} arises from 
  electron transitions due to disorder-assisted microwave absorption
  and emission. The {\it distribution-function mechanism} arises from
  microwave-induced changes in the electron distribution. However, the
  replacement of a smooth random potential by the unidirectional one,
  leads to different relative strengths of the two contributions to
  the photoconductivity. The distribution function mechanism dominates
  the photoconductivity in the direction of the static potential
  modulation, while both mechanisms contribute equally strongly to the
  photoconductivity in the perpendicular direction. Moreover, 
  the functional dependence of the negative
  photoconductivity on the microwave frequency is different for the
  two mechanisms, which may help to distinguish between them. In
  another marked difference from the case of smooth disorder, the
  unidirectionality of the static potential simplifies greatly the
  evaluation of the photoconductivities, which follow directly from
  Fermi's golden rule.
\end{abstract}

\maketitle

\section{Introduction}

Recent experiments \cite{Mani, Zudov, Yang, Dorozhkin, Willett} on a two-dimensional electron gas 
(2DEG) in weak magnetic fields under 
microwave irradiation have led to the unexpected discovery of regions in magnetic field where 
the longitudinal resistance is very close to zero. Unlike for quantized Hall states, the Hall 
resistance remains essentially classical and non-quantized for these novel ``zero-resistance 
states." These states occur near magnetic fields where, up to an additive constant, the microwave 
frequency $\omega$ is an integer multiple of the cyclotron frequency $\omega_c$. 

This discovery initiated a flurry of theoretical activity from which the following basic picture 
emerges. It has been argued \cite{Andreev, Bergeret} that under microwave irradiation, the microscopic 
diagonal conductivity can become negative. This would lead to
a (macroscopic) instability towards a current carrying state. 
Macroscopic resistance measurements on this state
show zero resistance because current can be made to flow through the sample by a rearrangement
of large current domains. Different microscopic mechanisms for a negative contribution to the 
microwave-induced photoconductivity have been proposed. One mechanism relies on disorder-assisted
absorption and emission of microwaves\cite{Durst, Shi, Lei, Vavilov} (see also Ref.\ \onlinecite{Ryzhii}).
Depending on the detuning $\Delta\omega
=\omega_c-\omega$, the displacement in real space associated with these processes is preferentially
in or against the direction of the applied $dc$ electric field. In an alternative mechanism, microwave 
absorption leads to a change in the electron distribution function, which can result in a negative 
photoconductivity.\cite{Dorozhkin, Dmitriev1, Dmitriev2} Detailed calculations within the self-consistent Born approximation
suggest that the latter mechanism is larger by a factor 
$\tau_{\rm in}/\tau_s^*$ where $\tau_{\rm in}$ is the inelastic relaxation time and $\tau_s^*$ denotes 
the single-particle elastic scattering time in a magnetic field. 

In the present paper, we study the microwave-induced photoconductivity within a model in which the 
2DEG is subjected to a unidirectional and static periodic potential. Our motivation for doing so 
is twofold. First, the study of periodically modulated 2DEGs in a perpendicular magnetic field
has led to the discovery of a number of interesting effects such as transport anisotropies
\cite{SmetWillett} and commensurability 
effects such as the Weiss
oscillations of the conductivity.\cite{Weiss, Gerhardts, Winkler, Beenakker} 
In addition, the periodic potential lifts the Landau level (LL) 
degeneracy. This allows one to exploit the familiar relation between momentum transfer and distance in real
space in high magnetic fields to compute the current by applying Fermi's golden rule. 
In this way, one finds in the absence of microwaves that scattering from the disorder
potential $U$ leads to a 
current 
\begin{eqnarray}
\label{current}
  j_x = {\pi e\over L_xL_y} \sum_{nn'} \sum_{kk'} (k'-k)\ell_B^2
   |\langle n'k' | U | nk\rangle|^2 [f^0_{nk}-f^0_{n'k'}]\delta(\epsilon_{nk}-\epsilon_{n'k'})
\end{eqnarray}
for an applied $dc$ electric field in the $x$-direction. Here, $f^0_{nk}$ is the Fermi-Dirac 
distribution function of the Landau level states $|nk\rangle$ which remain approximate 
eigenstates even in the presence of the periodic potential ($n$ is the LL index). The 
$\delta$-function involves the energies $\epsilon_{nk}$ of these states including the effect of both
periodic potential and $dc$ electric field.

It is evident from Eq.\ (\ref{current}) that the microwaves will affect the current in two ways.
(i) The joint effect of disorder and microwaves can give additional contributions to the transition 
matrix elements. This is the 
origin of the {\it displacement photocurrent} which relies on the displacements in real space associated
with disorder-assisted absorption and emission of microwaves. In more conventional terms, this contribution
can be associated with the effect of the microwaves on the collision integral in a kinetic equation.
(ii) The microwaves will also result in a redistribution of electrons, changing the electron distribution 
function $f_{nk}$ away from its equilibrium form $f_{nk}^0$. This {\it distribution-function contribution}
to the photocurrent will be important if inelastic relaxation is sufficiently slow. Our model
allows us to compute the various contributions to the photocurrent straight-forwardly within Fermi's
golden rule. 

For the parallel photocurrent (i.e., parallel to the wavevector of the static periodic modulation)
we find that the distribution-function mechanism gives a larger contribution than the 
displacement mechanism, by a factor $\tau_{\rm in}/\tau^*_s$. In addition, we find in this case that our 
results, with suitable identifications, are parametrically consistent with earlier results for disorder 
broadened Landau levels 
in the self-consistent Born approximation. \cite{Vavilov, Dmitriev2}
By contrast, we find a strong enhancement of 
the displacement mechanism for the 
perpendicular photocurrent so that in this case, both contributions are of the same order. 

This paper is organized as follows. In Sec.\ \ref{model}, we introduce the model and collect the relevant 
background material. In Sec.\ \ref{dark}, we compute the dark conductivity. The displacement mechanism
for the photocurrent is discussed in Sec.\ \ref{photoI}, while the distribution-function mechanism is
worked out in Sec.\ \ref{photoII}. Sec.\ \ref{photoII} also contains a discussion of the 
Weiss oscillations of the photocurrent. The polarization dependence is considered in Sec.\ \ref{apol}.
We summarize in Sec.\ \ref{conclusions}. Some technical
details are given in a number of appendices. In the remainder of this paper, we set $\hbar=1$.

\section{The model}
\label{model}

\subsection{Basics}

In this section, we specify our model and review some relevant background material. 
We consider a two-dimensional electron gas (2DEG) subject to a perpendicular magnetic field $B$ 
and a unidirectional static periodic potential 
\begin{equation}
   V({\bf r})={\tilde V}\cos(Qx)
\end{equation}
with period $a=2\pi/ Q$. The periodic potential which lifts the Landau level degeneracy, is assumed 
to be stronger than the residual disorder potential $U({\bf r})$. The disorder potential is characterized  
by zero average and variance
\begin{equation}
   \langle U({\bf r})U({\bf r}')\rangle = W({\bf r}-{\bf r}').
\end{equation}
For white-noise disorder, $W({\bf r}-{\bf r}')={1\over 2\pi\nu\tau} \delta({\bf r}-{\bf r}')$
with Fourier transform $\tilde W({\bf q})={1/ 2\pi\nu\tau}$.
Here, $\nu$ denotes the density of states at the Fermi energy in zero magnetic field and 
$\tau$ is the zero-field elastic scattering time. 
For smooth disorder potentials, the correlator $W({\bf r})$ falls off isotropically on the scale of the 
correlation length $\xi$ of the disorder potential ($\xi\gg \lambda_F$ for smooth disorder; $\lambda_F$ 
denotes the zero-field Fermi wavelength). 
We also note that the impurity average 
of the disorder matrix element between oscillator states $|nk\rangle$ for electrons in a magnetic 
field in the Landau gauge  is
\begin{equation}
\label{disordermatrixelement}
   |\langle nk' | U | nk \rangle|^2 = \int {d^2q\over (2\pi)^2}
       \delta_{q_y,k'-k} e^{-q^2\ell_B^2\over 2} [L_n({q^2\ell_B^2\over 2})]^2 \tilde W({\bf q}).
\end{equation}
Here, $n$ denotes the LL quantum number and $k$ the momentum in the $y$ direction. $L_n(x)$ denotes the 
Laguerre polynomial and $\ell_B=(\hbar/eB)^{1/2}$ the magnetic length.

In this paper, we focus on the regime of high Landau levels so that $\lambda_F\ll \ell_B\ll R_c$.
(Here, $R_c$ denotes the cyclotron radius.) We assume that the period $a$ of the modulation 
satisfies the condition
\begin{equation}
\label{condition:white}
   \lambda_F \ll a \ll R_c.
\end{equation}
This is essentially a technical condition, which simplifies some of the calculations. 
For smooth disorder, we assume, in addition, that the
correlation length $\xi$ of the disorder potential satisfies the inequality
\begin{equation}
\label{condition:smooth}
   \lambda_F \ll \xi \ll \ell_B^2/a.
\end{equation}
Here, the first inequality reflects the fact that the disorder is smooth, while the second inequality
ensures that the typical jump in real space of length $\ell_B^2/\xi$ associated with a disorder scattering event is 
large compared to the period $a$ of the periodic modulation. 

The 2DEG is irradiated by microwaves described by the electric potential 
\begin{equation}
\phi({\bf r},t) = -{e\over 2} {\bf r} ({\bf E}^*e^{i\omega t} + {\bf E}
e^{-i\omega t}) = \phi_+ e^{-i\omega t} + \phi_-
e^{i\omega t},
\end{equation}
where $\phi_+ = [\phi_-]^* =- e{\bf Er}/2$. The complex vector ${\bf E}$ contains
both strength and polarization of the microwaves.

In the absence of disorder and microwaves, and for sufficiently weak periodic potential, 
the single-particle spectrum of the electrons can be obtained by treating the 
periodic potential perturbatively. Starting with the oscillator states $|nk\rangle$, one 
obtains
\begin{equation}
   \epsilon^0_{nk} \simeq \omega_c (n+{1\over 2}) + V_{n}\cos(Qk\ell_B^2).
\end{equation}
The amplitude $V_n$ is given
by
\begin{equation}
\label{Weissosc}
   V_n = {\tilde V} e^{-Q^2\ell_B^2/4}L_n(Q^2\ell_B^2/2).
\end{equation}
In the limit of high Landau levels, $V_n$ can be approximated as $V_n \simeq {\tilde V}J_0(QR_c)$ and 
thus exhibits slow oscillations with period 
$k_Fa\gg 1$ as a function of LL index $n$. (The LL index $n$ enters via the cyclotron radius.) 
This also implies oscillations of $V_n$ as function of the magnetic field.
It is these oscillations of $V_n$ which are responsible
for the Weiss oscillations \cite{Gerhardts} of the conductivity.

If in addition, a $dc$ electric field $E_{\rm dc}$  is applied in the 
$x$-direction, the eigenenergies take the form
\begin{equation}
      \epsilon_{nk} \simeq \epsilon_{nk}^0-eE_{\rm dc} k\ell_B^2. 
\end{equation}
It is useful to define the density of states (DOS) of a periodic potential broadenend LL by
\begin{equation}
   \nu^*(\epsilon) = \nu^* \tilde\nu^*(\epsilon)
\end{equation}
with the density of states at the band center
\begin{equation}
  \nu^*={1\over 2\pi\ell_B^2} {1\over \pi V_n}
\end{equation}
and the normalized density of states
\begin{equation}
\label{DOS}
   \tilde\nu^*(\epsilon) = {1\over \sqrt{1-[(\epsilon-E_n)/V_n]^2}}.
\end{equation}
Here, $n$ and $\epsilon$ satisfy $|\epsilon-E_n|<V_n$ (with the LL energy $E_n=\omega_c(n+1/2)$ ).  
Note that the DOS can also be expressed as $\nu^*\sim\nu (\omega_c/V_n)$,
reflecting the increased density of states due to the Landau quantization. 

\subsection{Kinetic equation}

We now turn to setting up the kinetic equation for
the non-equilibrium electronic distribution 
function $f_{nk}$ which describes the occupation of the LL oscillator eigenstates $|nk\rangle$.
These occupations change due to disorder scattering, disorder-assisted microwave absorption and emission,
as well as inelastic relaxation which we include within the relaxation-time approximation. 
Note that in principle this distribution function depends also on the spatial coordinate $y$. However,
it will be sufficient throughout this work to consider distribution functions which are uniform 
in the $y$ direction. (The dependence on $x$, on the other hand, 
is included, as the momentum $k$ also plays the role of a position in the $x$-direction.)

If the $dc$ electric field points in the $x$-direction, the kinetic equation takes the form
\begin{eqnarray}
   {\partial f_{nk}\over \partial t} = \left( {\partial f_{nk}\over \partial t} \right)_{\rm dis}
       + \left( {\partial f_{nk}\over \partial t} \right)_{\rm mw} - { f_{\rm nk} - f^0_{\rm nk} \over
      \tau_{\rm in}}.
\end{eqnarray}
In principle, there should also be a term which describes the drift in the $y$-direction induced by the 
$dc$ electric field. However, this term has no consequences when considering distribution functions which
are independent of $y$. In the last term on the right-hand side, $f_{nk}^0$ denotes the equilibrium
Fermi-Dirac distribution and $\tau_{\rm in}$ denotes a phenomenological inelastic relaxation rate. 
The collision integral for disorder scattering is explicitly given by
\begin{eqnarray}
\label{coll-disorder}
   \left( {\partial f_{nk}\over \partial t} \right)_{\rm dis} 
    =\sum_{n'k'} 2\pi |\langle n'k' | U | nk \rangle |^2 [ f_{n'k'} - f_{nk} ]
        \delta ( \epsilon_{nk} - \epsilon_{n'k'} ).
\end{eqnarray}
The collision integral for disorder-assisted microwave absorption and emission is
\begin{eqnarray}
\label{coll-micro}
   \left( {\partial f_{nk}\over \partial t} \right)_{\rm mw} 
    =\sum_{n'k'}\sum_{\sigma=\pm} 2\pi |\langle n'k' | T_\sigma | nk \rangle |^2 [ f_{n'k'} - f_{nk} ]
        \delta ( \epsilon_{nk} - \epsilon_{n'k'} +\sigma\omega ).
\end{eqnarray}
The precise nature of the operator $T_\sigma$ will be given in Eq.\ (\ref{microwavematrixelement}) below. 
Note that these collision integrals involve the electron energies including the effects of the 
$dc$ electric field. 

If the $dc$ electric field points along the $y$-direction, we can no longer include it in the 
eigenenergies. Instead, it enters the kinetic equation through
an additional term describing the associated drift in the $x$-direction,
\begin{eqnarray}
   {\partial f_{nk}\over \partial t} = - eE_{\rm dc}{\partial f_{nk}\over\partial k}  +
     \left( {\partial f_{nk}\over \partial t} \right)_{\rm dis}
       + \left( {\partial f_{nk}\over \partial t} \right)_{\rm mw} - { f_{\rm nk} - f^0_{\rm nk} \over
      \tau_{\rm in}}.
\label{kineticy}
\end{eqnarray}
The collision integrals are given by the expressions in Eqs.\ (\ref{coll-disorder}) and (\ref{coll-micro}) 
with the energies in the $\delta$-functions taken in the absence of the $dc$ electric field. 

We close this section with a calculation of the elastic scattering rate $1/\tau^*$ in high magnetic fields.
The motivation for doing this is twofold. First, $\tau^*$ is a natural parameter in terms of 
which to write our final results for the conductivity. On a more technical note, 
computing $\tau^*$ gives us the 
opportunity to introduce a convenient way of dealing with integrals involving Laguerre polynomials,
which will be used repeatedly throughout this paper. From the collision integral for elastic disorder 
scattering, we obtain
\begin{equation}
  {1\over\tau^*(\epsilon)} 
    =\sum_{n'k'} 2\pi |\langle n'k' | U | nk \rangle |^2   \delta ( \epsilon - \epsilon_{n'k'} )
\end{equation}
with $\epsilon=\epsilon_{nk}$. Noting that $n=n'$ and inserting the expression (\ref{disordermatrixelement}) 
for the matrix element, we obtain
\begin{equation}
\label{Laguerreintegral}
  {1\over\tau^*(\epsilon)} 
    =  2\pi \int {d^2q\over (2\pi)^2}
        e^{-q^2\ell_B^2\over 2} [L_n({q^2\ell_B^2\over 2})]^2 \tilde W({\bf q})
   \delta ( \epsilon - \epsilon_{nk+q_y} ).
\end{equation}
The Laguerre-polynomial factor arising from the matrix elements of the disorder potential decays as 
a function of $q\ell_B^2$ on the scale of the cyclotron radius $R_c$, in addition to fast oscillations
on the scale of the zero-field Fermi wavelength $\lambda_F$. On the other hand, the argument of the 
$\delta$-function changes with $k\ell_B^2$ on the scale of the period $a$ of the periodic potential.
Thus, for white-noise disorder and in the 
limit $\lambda_F \ll a \ll R_c$, we can average the $\delta$-function separately over $q_y$. Using the identity
\begin{equation}
   \langle \delta(\epsilon-\epsilon^0_{nk'})\rangle_{k'}= 2\pi\ell_B^2 \nu^*(\epsilon)
\end{equation}
and performing the remaining integral over the Laguerre polynomial, 
\begin{equation}
   \int {d^2q\over (2\pi)^2}
        e^{-q^2\ell_B^2\over 2} [L_n({q^2\ell_B^2\over 2})]^2={1\over 2\pi\ell_B^2},
\end{equation}
we find the result
\begin{equation}
\label{zerofinite}
  {1\over\tau^*(\epsilon)} 
    =  {1 \over \tau} {\nu^*(\epsilon)\over\nu}.
\end{equation}
In the following, we will also use the notation $\tau^*=\tau^*(\epsilon=E_n)$, i.e.\
\begin{equation}
   \tau^*=\tau {\pi V_n\over \omega_c}.
\end{equation} 
This result reflects the increased density of final states in the limit of well-separated Landau levels.

For smooth disorder, we need to distinguish between the single-particle scattering time and the 
transport scattering time. Their zero-field values $\tau_s$ and $\tau_{\rm tr}$ 
are related to the
finite field values $\tau_s^*(\epsilon)$ and $\tau_{\rm tr}^*(\epsilon)$ in analogy to 
Eq.\ (\ref{zerofinite}), i.e., 
$\tau_s^*(\epsilon) = \tau_s \nu/\nu^*(\epsilon)$ and $\tau_{\rm tr}^*(\epsilon) = \tau_{\rm tr} 
\nu/\nu^*(\epsilon)$. Some details of the calculation are given in App.\ \ref{smooth}.

\section{Dark conductivity}
\label{dark}

\subsection{Conductivity $\sigma_{xx}$ along the modulation direction}

In this section, we compute the dark conductivity, i.e., the conductivity in the 
absence of microwaves. We start with the situation in which the $dc$ electric field is applied in the 
$x$-direction, i.e.\ parallel to the wavevector of the static periodic modulation. We assume that the 
$dc$ electric field is sufficiently weak so that heating effects can be ignored. In this case, the distribution 
function remains in equilibrium, $f_{nk}=f_{nk}^0$, and the system responds to the $dc$ electric field with a 
current in the $x$-direction. 

This current can be expressed by counting the number of disorder scattering events that take an electron
from a state $k$, localized in the $x$-direction at $k\ell_B^2$ to the left of an imaginary line $x_0$ 
parallel to the $y$-axis, to a state $k'$, localized at $k'\ell_B^2$ to the right of this imaginary line, 
and vice versa. Due to current conservation, the current is independent of the particular choice of $x_0$
and it turns out to be useful to average over all possible $x_0$. This results in the expression 
\begin{eqnarray}
 j_x = {e\over L_y} \int_{-L_x/2}^{L_x/2} {dx_0\over L_x}\sum_{nn'}
   \sum_{k<{x_0\over\ell_B^2}}\sum_{k'>{x_0\over\ell_B^2}}
   2\pi|\langle n'k' | U | nk\rangle|^2[f^0_{nk}-f^0_{n'k'}]
  \delta(\epsilon_{nk}-\epsilon_{n'k'})
\end{eqnarray}
for the current in the $x$-direction. Performing the integral over $x_0$ gives
\begin{eqnarray}
     j_x = {\pi e\over L_xL_y} \sum_{nn'} \sum_{kk'} (k'-k)\ell_B^2
   |\langle n'k' | U | nk\rangle|^2
  [f^0_{nk}-f^0_{n'k'}]
  \delta(\epsilon_{nk}-\epsilon_{n'k'}).
\end{eqnarray}
Inserting the explicit expression (\ref{disordermatrixelement}) for the disorder-averaged matrix 
element and performing the sum over $k'$, one obtains
\begin{eqnarray}
   j_x  ={\pi e\over L_xL_y} \sum_{nk}
   \int {d^2q\over (2\pi)^2} q_y\ell_B^2 e^{-q^2\ell_B^2\over 2} [L_n({q^2\ell_B^2\over 2})]^2
     \tilde W({\bf q}) [f^0_{nk}-f^0_{nk+q_y}] 
\delta(\epsilon^0_{nk}-\epsilon^0_{nk+q_y} + eE_{\rm dc} q_y\ell_B^2).
\end{eqnarray}
Expanding to linear order in the $dc$ electric field, one obtains for the conductivity
\begin{eqnarray}
\label{prelsigmadarkxx}
   \sigma_{xx} = {\pi e^2\over L_xL_y} \sum_{nk}
   \left(-{\partial f^0_{nk}\over \partial \epsilon^0_{nk}}\right)
   \int {d^2q\over (2\pi)^2} (q_y\ell_B^2)^2 e^{-q^2\ell_B^2\over 2} [L_n({q^2\ell_B^2\over 2})]^2
   \tilde W({\bf q}) \delta(\epsilon^0_{nk}-\epsilon^0_{nk+q_y}).
\end{eqnarray}
For white-noise disorder and $\lambda_F \ll a\ll R_c$, the Laguerre-polynomial integral can be computed  
in analogy with the 
evaluation of Eq.\ (\ref{Laguerreintegral}) above. 
Using
\begin{equation}
   \int {d^2q\over (2\pi)^2} (q_y\ell_B^2)^2 e^{-q^2\ell_B^2\over 2} [L_n({q^2\ell_B^2\over 2})]^2
      = {N\over \pi},
\end{equation}
this yields 
\begin{eqnarray}
\label{sigmadarkxxprel}
  \sigma_{xx} = {e^2N\over L_xL_y}{1 \over 2\pi \nu \tau} \sum_{nk}
   \left(-{\partial f^0_{nk}\over \partial \epsilon^0_{nk}}\right)
    2\pi\ell_B^2\nu^*(\epsilon^0_{k}).
\end{eqnarray}
Expressing the sum by an integral involving the density of states, we can cast this result in the
final form
\begin{eqnarray}
\label{sigmadarkxx}
    \sigma_{xx} = \int d\epsilon \left(-{\partial f(\epsilon)\over \partial \epsilon}\right)
      \sigma_{xx}(\epsilon)
\end{eqnarray}
in terms of  
\begin{eqnarray}
\label{sigmaepsilondarkxx}
    \sigma_{xx}(\epsilon) = e^2 \left({R_c^2\over 2\tau_{\rm tr}^*(\epsilon)}\right)\nu^*(\epsilon).
\end{eqnarray}
This equation is written such that it includes both types of disorder. For white-noise disorder 
$\tau_{\rm tr}^*=\tau_s^*=\tau^*$, while for smooth disorder $\tau_{\rm tr}^*\neq \tau_s^*$.
The derivation of the result for smooth disorder is sketched in App.\ \ref{smooth}.
 
This result for the dark conductivity can be interpreted as follows. The bare rate for disorder scattering
is $1/\tau_s^*$, where each scattering event is associated with a momentum transfer $1/\xi$. This 
momentum transfer translates into a jump of magnitude $\ell_B^2/\xi$ in real space so that the electron
diffuses in the $x$-direction with a diffusion constant $D_{xx}\sim(\ell_B^2/\xi)^2/\tau_s^*$. 
Alternatively, this diffusion constant can be written in terms of the transport time as 
$D_{xx}=R_c^2/2\tau^*_{\rm tr}$ (using that $\tau_{\rm tr}^*/\tau_{\rm s}^* \sim (k_F\xi)^2$).
By the Einstein relation, this diffusion constant translates into the conductivity given in Eq.\ (\ref{sigmaepsilondarkxx}). The conductivity (\ref{sigmaepsilondarkxx}) can also be expressed in terms
of the zero-B conductivity $\sigma_{xx}(B=0)$ as  $\sigma_{xx}=\sigma_{xx}(B=0)/(\omega_c \tau^*_{\rm tr})^2$.
We also note that $\sigma_{xx} \sim 1/V_n^2$ so that the oscillations of $V$ with magnetic field $B$
(cf.\ Eq.\ (\ref{Weissosc}) above) lead to Weiss oscillations of the conductivity, in agreement with previous 
results.\cite{Gerhardts}

The energy integral in Eq.\ (\ref{sigmadarkxx}) is formally logarithmically divergent due to the
square-root singularity of the density of states $\nu^*(\epsilon)$ at the band edge. This singularity is
cut off by smearing of the band edge by disorder or by the applied $dc$ electric field, when the latter 
is kept beyond linear order.

\subsection{Conductivity $\sigma_{yy}$ perpendicular to the modulation direction}

An applied $dc$ electric field in the $y$-direction leads to a non-equilibrium distribution
function $f_{nk}$ due to the drift term in the kinetic equation (\ref{kineticy}). In the absence of microwaves, 
linearizing the stationary kinetic equation in the applied $dc$ electric field yields
\begin{eqnarray}
eE_{\rm dc}{\partial f^0_{nk}\over\partial k}
  = 2\pi  
\int {d^2q\over (2\pi)^2} e^{-q^2\ell_B^2\over 2} 
[L_n({q^2\ell_B^2\over 2})]^2 \tilde W({\bf q}) [ \delta f_{nk+q_y} - \delta f_{nk} ]
\, \delta(\epsilon^0_{nk}-\epsilon^0_{nk+q_y}).
\end{eqnarray}
Here, $ \delta f_{nk}= f_{nk}-f^0_{nk} $ denotes the deviation from the equilibrium distribution function, 
and we have neglected inelastic processes relative to elastic disorder scattering. Due to the 
periodicity in the $x$-direction, $\delta f_{nk}=\delta f_{nk+a/\ell_B^2}$. Moreover, 
if $k$ and $k+q_y$ are two momenta with the same energy $\epsilon_{nk}^0$, but opposite signs of the 
derivative $\partial \epsilon_{nk}^0/\partial k$, then $\delta f_{nk}= - \delta f_{n k+q_y}$. 
Using that for white-noise disorder and $ \lambda_F\ll a \ll R_c$, we can split the $q$-integration as in 
the evaluation of (\ref{Laguerreintegral}) and obtain  
\begin{equation}
\delta f_{nk}= - eE_{\rm dc}\tau^*(\epsilon^0_{nk}) 
{\partial f^0_{nk}\over\partial k}.
\end{equation} 
In terms of the distribution function, the current in the $y$-direction is given by 
\begin{equation}
j_y=e \frac{1}{L_x L_y} \sum_{n k} {\partial \epsilon^0_{nk}\over\partial k}
\; \delta f_{nk}.
\end{equation} 
Inserting the expression for the distribution function gives for the conductivity
\begin{equation}
\label{sigmadarkyyprel}
   \sigma_{yy}=-{e^2 \over L_xL_y} \sum_{nk}
      \left(\partial\epsilon^0_{nk}\over \partial k\right)^2 \tau^*(\epsilon^0_{nk})
      {\partial f^0_{nk}\over\partial \epsilon^0_{nk}}. 
\end{equation}
Expressing the sum over $nk$ as an energy integral involving the DOS, we obtain
\begin{eqnarray}
 \sigma_{yy}= 
  \int d\epsilon \left(-{\partial f^0(\epsilon)\over \partial \epsilon}\right) \sigma_{yy}(\epsilon)
\end{eqnarray}
with 
\begin{eqnarray}
\label{sigmadarkyy}
\sigma_{yy}(\epsilon) =  e^2 \left([v_y(\epsilon)]^2\tau_s^*(\epsilon)\right)
   \nu^*(\epsilon)    .
\end{eqnarray}
As above for $\sigma_{xx}$, this result is written such that it includes both the case of 
white-noise and of smooth disorder. The derivation for the case of smooth disorder is 
sketched in App.\ \ref{smooth}. We have defined the drift velocity
\begin{equation}
   |v_y(\epsilon)| =  \left|\partial\epsilon^0_{nk}\over \partial k\right| =
      {1\over\pi a \nu^*(\epsilon) }  
\end{equation}
in the $y$-direction, induced by the periodic modulation. 

The result (\ref{sigmadarkyy}) can be 
interpreted as follows. With respect to the motion in the $y$-direction, a partially filled LL 
consists effectively of a set of two ``internal edge channels" parallel to the 
$y$-axis per period $a$. Neighboring channels flow in opposite directions so that disorder
scattering randomizes the direction of the motion in the $y$-direction after time $\tau_s^*$. 
The factor $D_{yy}=v_y^2 \tau_s^*$ can thus be interpreted as the diffusion constant of the resulting 
diffusion process.
We note that unlike $\sigma_{xx}(\epsilon)$, the conductivity $\sigma_{yy}(\epsilon)$ remains 
finite at the band edge. The anisotropy $\sigma_{yy}/\sigma_{xx}$ of the dark conductivity
is thus of order $(v_y\tau_s^*/R_c)^2(k_F \xi)^2$, where both factors are larger than unity. 
The dark conductivity $\sigma_{yy}$ depends on the modulation-induced LL broadening as
$\sigma_{xx} \sim V_n^2$, so that the Weiss oscillations in $\sigma_{yy}$ are phase shifted by $\pi$ 
relative to the oscillations in $\sigma_{xx}$, in agreement with standard 
results.\cite{Gerhardts}

The $dc$ electric field also leads to heating of the electron system. The
characteristic field $E_{\rm dc}^*$ where this becomes relevant, can be estimated as follows. 
The $dc$ electric field causes a drift in the $x$-direction with drift velocity $(E_{\rm dc}/B)$,
changing the potential energy of the electron by $(V/a)(E_{\rm dc}/B)\tau_s^*$. This gives rise to a 
diffusion constant in energy of $D_\epsilon\sim(V/a)^2(E_{\rm dc}/B)^2\tau_s^*$. Heating can be 
neglected as long as the typical energy change $(D_\epsilon\tau_{\rm in})^{1/2}$ is small compared 
to $V$. This gives the condition
\begin{equation}
   E_{\rm dc} \ll E_{\rm dc}^*={Ba\over 2\pi\sqrt{\tau_{\rm in}\tau_s^*}}
\end{equation}
for the $dc$ electric field. It is only for these electric fields that the result (\ref{sigmadarkyy})
is valid. A more formal derivation of this result is given in App.\ \ref{heating}.

For larger $dc$ electric fields $E_{\rm dc}\gg E_{\rm dc}^*$, the effect of heating needs to be taken into 
account. Following the arguments given in App.\ \ref{heating}, one expects that the conductivity is suppressed 
by heating effects and behaves in magnitude as 
\begin{equation}
\label{heatingsigmayy}
   \sigma_{yy} \sim  \left(E^*_{\rm dc}\over E_{\rm dc}\right)^2 e^2D_{yy} \nu^*.
\end{equation}
The reason for this suppression is that heating reduces the $k$-dependence of the distribution function. 

\section{Displacement photocurrent}
\label{photoI}

\subsection{$t$-matrix elements}

The microwaves lead to additional contributions to the transition matrix element between 
LL oscillator states which enters into the current expression (\ref{current}). Direct microwave 
absorption or emission does not contribute to the current, because the microwaves do not 
transfer momentum to the electrons so that such processes are not associated with displacements
in real space. In addition, such processes occur only for $\omega=\omega_c$. On the other hand, 
disorder-assisted microwave absorption and emission is associated with displacements in real 
space of the order of $R_c$ ($\ell_B^2/\xi$ for smooth disorder). 
This process is allowed for microwave frequencies away 
from $\omega_c$. In this section, we compute the contribution of this displacement mechanism
to the photoconductivity within our model. 

The transition rate between LL oscillator states involves the $t$-matrix
\begin{equation}
    T = (U+\phi) + (U+\phi)G_0(U+\phi) + \ldots ,
\end{equation}
where $G_0$ denotes the retarded Green function of the unperturbed system. The dark
conductivity, computed in the previous section, follows in the approximation
$T\simeq U$. Disorder-assisted microwave absorption $T_+$ and emission $T_-$ is 
given by
\begin{equation}
\label{microwavematrixelement}
    T_\pm = [UG_0\phi_\pm + \phi_\pm G_0 U ].
\end{equation}
Note that $T_+$ and $T_-$ contribute incoherently. Assuming that $\omega$ couples only neighboring LLs
and that the microwaves are linearly polarized in the $x$-direction,
the corresponding matrix elements between LL oscillator states are 
\begin{eqnarray}
   \langle n\pm 1 k' | T_\pm | n k \rangle 
     &=& \pm \left({eER_c\over 4\Delta\omega}\right) [ \langle n \pm 1 k' |U|n \pm 1 k\rangle
      -\langle n k'|U|n k\rangle ],
\end{eqnarray}
where we used
\begin{eqnarray}
    G_{0,n\pm 1 k}(\epsilon_{nk}\pm\omega) &=& {1\over \epsilon_{nk} \pm \omega - \epsilon_{n\pm 1 k}}
      = \mp {1\over \Delta\omega}
      \nonumber \\
    G_{0,n k'}(\epsilon_{nk}) &=& {1\over \epsilon_{nk} - \epsilon_{nk'} }= \pm {1\over \Delta\omega}.
\end{eqnarray}
Using the disorder matrix elements and neglecting corrections of order $1/n$, one finds
\begin{eqnarray}
\label{photoImatrixelement}
     |\langle n\pm 1 k' | T_\pm | n k \rangle|^2
       \simeq \left({eER_c\over 4\Delta\omega}\right)^2 
       \int {d^2q\over (2\pi)^2} \,\delta_{q_y,k'-k} e^{-{q^2\ell_B^2\over 2}} [ L_{n+ 1} ({q^2\ell_B^2\over 2})
        -L_n({q^2\ell_B^2\over 2})]^2 \tilde W({\bf q}).
\end{eqnarray}
Thus in this approximation, the matrix elements are identical for absorption and emission and depend only on
the absolute value of $k-k'$. It is worthwhile to point out that the difference between the two Laguerre
polynomials reflects the fact that disorder-assisted microwave absorption involves a coherent sum of two 
processes: In one process, a microwave photon is first absorbed, resulting in a transition from the $n$th
to the $(n+1)$th LL, with disorder subsequently inducing a transition between states in the $(n+1)$th
LL. In the second process, disorder first leads to a transition between states in the $n$th LL
with a subsequent absorption of a microwave photon. The divergence of the matrix element for 
$\Delta\omega\to 0$ is an artefact of low-order perturbation theory in the disorder potential $U$. 
In a more accurate treatment, this divergence would be removed by disorder broadening.

\subsection{Displacement photocurrent $j_x$ along the modulation direction}

In this section, we compute the displacement contribution to the photocurrent in the 
modulation direction. 
The current in the $x$-direction can now be computed in terms of Fermi's golden rule 
in the same manner as for the dark current in Sec.\ \ref{dark}. In this way, one obtains
\begin{eqnarray}
   j_x^{\rm photo\,I} = {2\pi e \over L_xL_y} \sum_{n} \sum_{kk'} (k'-k)\ell_B^2
    |\langle n+1 k' | T_+ | n k \rangle|^2
   [f^0_{nk} - f^0_{n+1k'}] \delta(\epsilon_{nk} -\epsilon_{n+1k'} + \omega).
\end{eqnarray}
Here, we assume that the microwaves frequency $\omega>0$ is such that it couples neighboring LLs. 

In order not to complicate the calculations unnecessarily, we will consider temperatures
$T\gg V$. In this regime, the temperature smearing is over an energy range large 
compared to the LL width and the distribution function depends only on the LL index $n$, but not
on the momentum $k$. Inserting the explicit expression (\ref{photoImatrixelement}) 
for the $t$-matrix element, we obtain
\begin{eqnarray}
   j_x^{\rm photo\, I}
    &=& {2\pi\ell_B^2 e \over L_xL_y}\left({eER_c\over 4\Delta\omega}\right)^2 
         \sum_n [f_n - f_{n+1}]
         \nonumber\\
         &&\,\,\,\,\,\,\times\sum_{k}
       \int {d^2q\over (2\pi)^2} q_y e^{-{q^2\ell_B^2\over 2}} [ L_{n + 1} ({q^2\ell_B^2\over 2})
        -L_n({q^2\ell_B^2\over 2})]^2 \tilde W({\bf q})
         \delta(\epsilon_{n+1 k} -\epsilon_{nk+q_y} -\omega).
\end{eqnarray}
The $k$-summation gives
\begin{eqnarray}
  \sum_k \delta(\epsilon_{n+1 k} -\epsilon_{nk+q_y} -\omega)
     = {L_xL_y\over 2\pi \ell_B^2 2\pi V_n} {1\over [\sin^2 {Qq_y\over 2} - ({\Delta\tilde\omega\over
       2V_n})^2]^{1/2}},
\end{eqnarray}
where we introduced $\Delta\tilde\omega = \Delta \omega - eE_{\rm dc} q_y\ell_B^2$.
Thus, we obtain for the current
\begin{eqnarray}
\label{photoIxxintermediate}
   j_x^{\rm photo\, I}
    &=& {e \over 2\pi \ell_B^2}\left({eER_c\over 4\Delta\omega}\right)^2 
         \sum_n [f_n - f_{n+1}]
         \nonumber\\
         &&\,\,\,\,\,\,\times
       \int {d^2q\over (2\pi)^2} q_y\ell_B^2 e^{-{q^2\ell_B^2\over 2}} [ L_{n + 1} ({q^2\ell_B^2\over 2})
        -L_n({q^2\ell_B^2\over 2})]^2 {\tilde W({\bf q})\over V_n[\sin^2 {Qq_y\over 2} - ({\Delta\tilde\omega\over
       2V_n})^2]^{1/2}},
\label{jumprate}
\end{eqnarray}
where the integral is only over the region where the square root in
the denominator is real. It is useful to interpret the various factors in this expression. 
It consists of a charge density per LL ${e/ 2\pi\ell_B^2}$, and a rate (per LL) for jumps in the $x$-direction 
with lengths between $q_y\ell_B^2$ and $(q_y+dq_y)\ell_B^2$, multiplied by the jump lengths $q_y\ell_B^2$ 
Finally, the expression is integrated over all jump lengths $q_y$ and summed over all LLs.

\begin{figure}
 \psfrag{A1}[][][.7]{$\hspace*{.2cm} A_1 $ }
 \psfrag{B1/ln(V/B)}[][][.8] {$\hspace*{.5cm} B_1/\ln(V/\Delta) $}
 \psfrag{x}[][][.7] {\hspace*{-.15cm} $\Delta\omega/2V$ }
\epsfig{file=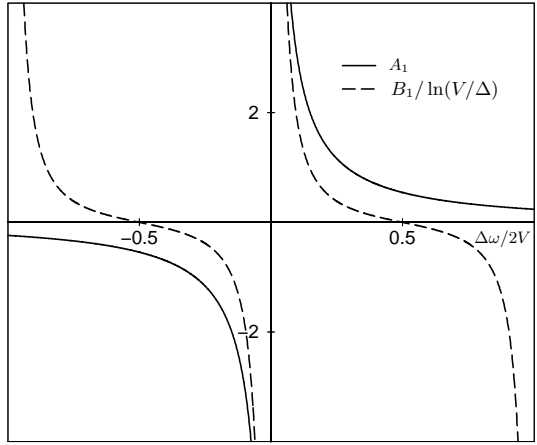,width=7cm}
\caption{The functions $A_1(\Delta\omega/2V_N)$ (full line) and $B_1(\Delta\omega/2V_N)/\ln(V_N/\Delta)$ (dashed line),
describing the dependence of the parallel photoconductivity on the microwave frequency, 
cf.\ Eqs.\ (\ref{A1}) and (\ref{B11}).}
\label{fig1}
\end{figure}

The sum over LLs $n$ is trivial and for white-noise disorder, the integral over $q$ can again be decoupled 
for $ \lambda_F \ll a\ll R_c$. Expanding to linear order in the $dc$ electric field and using the integral
\begin{equation}
 \int {d^2q\over (2\pi)^2} (q_y\ell_B^2)^2 e^{-{q^2\ell_B^2\over 2}} [ L_{n + 1} ({q^2\ell_B^2\over 2})
        -L_n({q^2\ell_B^2\over 2})]^2 = {3n\over \pi},
\end{equation}
we obtain the linear-response 
conductivity   
\begin{eqnarray}
\label{sigmaxxphotoIfinal}
  \sigma_{xx}^{\rm photo\,I} =  \left[e^2D_{xx}\nu^* \right] {\tau_s^*\over\tau^*_{\rm tr}}  
    \left({eER_c\over 4\Delta\omega}\right)^2 A_1(\Delta\omega/2V_N)
\end{eqnarray}
with the function (cf.\ Fig.\ \ref{fig1})
\begin{eqnarray}
\label{A1}
    A_1(x)=-{3\over \pi^2}\frac{\partial}{\partial x} 
  K\left(\sqrt{1- x^2}\right), 
\end{eqnarray} 
expressed in terms of the complete elliptic function $K$. 
Note that the first factor in Eq.\ (\ref{sigmaxxphotoIfinal})
is just the dark conductivity $\sigma_{xx}$ [cf.\ Eq.\ (\ref{sigmaepsilondarkxx})]
and that the result includes the 
case of smooth disorder. (Strictly speaking, the result for smooth disorder is valid only up to a numerical
prefactor that depends on the precise nature of the smooth disorder potential, cf.\ App.\ \ref{smooth} for 
details.)
The behavior of this displacement photocurrent 
for $\Delta\omega\ll 2V_N$ and $\Delta\omega \sim 2V_N$ follows from the asymptotic expressions
\begin{equation} 
  K(\sqrt{1- (\Delta \omega /2 V_N)^2})  \simeq \left\{ 
    \begin{array}{cc} -\ln(|\Delta \omega| /8 V_N) & |\Delta \omega /2V_N| \ll 1 \\
    {\pi\over 4} (1+{\alpha\over 4}) & \alpha = 1- ({\Delta\omega\over 2V_N})^2 \ll 1
    \end{array} \right. .
\end{equation}  
This implies that $ A_1(\Delta\omega/2V_N) $ remains finite for $ |\Delta \omega/2V_N| \to 1 $ 
(cf.\ Fig.\ \ref{fig3}) and 
is proportional to $ 1/\Delta \omega $ for small $ |\Delta \omega/2V_N|$.
The sign of the displacement photocurrent is given by ${\rm sgn} 
(\omega_c-\omega)$, similar to previous work on disorder-broadenend LLs.\cite{Durst,Vavilov}

{\center
\begin{figure}
\psset{xunit=0.01,yunit=1cm}
\hspace*{-5.cm}
\vspace*{-0.3cm}
\scalebox{0.7}{
\begin{pspicture}(-3,-1.0)(8,5)
\psline[linewidth=0.5pt,arrowsize=0.15]{->}(-40,-0.5)(-40,4.6)
\rput(-0.5,2.0){\large \boldmath $ \epsilon $ } 
\psline[linewidth=1.5pt]{->}(321,0.24)(321,3.7)
\psline[linewidth=1.5pt,linestyle=dashed]{<->}(193,3.7)(323,3.7)(476,3.7)  

\psplot[plotstyle=curve,linewidth=1.5pt]{50}{720}{x -50 add  sin 0.6 
mul 1 -0.0006 x mul add add}  
\psplot[plotstyle=curve,linewidth=1.5pt ]{50}{720}{x -50 add  sin 0.6 
mul 3.4 -0.0006 x mul add add}
 
\psplot[plotstyle=curve,linewidth=0.7pt,linestyle=dotted]
{50}{720}{x -50 add  sin 0 mul 1 -0.0006 x mul add add}  
\psplot[plotstyle=curve,linewidth=0.7pt,linestyle=dotted]{50}{720}
{x -50 add  sin 0 mul 3.4 -0.0006 x mul add add}

\rput(258,4.0){\boldmath $ U $} 
\rput(383,4.0){\boldmath $ U $}
\rput(352,1.7){\large \boldmath $ \phi $ }   
\rput(740,0.3){ n  }
\rput(740,2.8){ n+1  }
\end{pspicture}}
\caption{Illustration of disorder-assisted microwave absorption for $\Delta\omega\simeq 2V_N$. 
In the $x$-direction, the Landau levels are modulated by the periodic potential and tilted by the 
$dc$ field. 
The square-root 
singularity of the Landau level DOS (\ref{DOS}) at the band edge leads to the singular behavior 
of the photocurrent
for these detunings $\Delta\omega$, cf.\ Fig.\ \ref{fig1}.  }
\label{fig3}
\end{figure} 
}
  
The magnitude of the displacement contribution (\ref{sigmaxxphotoIfinal}) to the photoconductivity can be understood 
as follows. The bare rate of disorder-induced microwave absorption is $(1/\tau^*_{\rm s})(eE
R_c/\Delta\omega)^2$, where the second factor is the dipole coupling of the microwave field for Landau states 
divided by the relevant energy denominator of the intermediate state. Each of these scattering events is 
associated with a jump in real space of the order of $\ell_B^2/\xi$, resulting in an effective
diffusion constant $D_{xx} (eE R_c/\Delta\omega)^2$. An additional factor $(\tau^*_s/
\tau_{\rm tr}^*)$ arises because of the partially destructive interference of the two contributions 
which were discussed
below Eq.\ (\ref{photoImatrixelement}). This interference leads to the difference of Laguerre 
polynomials in Eq.\ (\ref{photoIxxintermediate}) which introduces an additional factor $(q/k_F)^2$ into the 
integral. This factor is of order $1/(k_F\xi)^2\sim (\tau^*_s/\tau_{\rm tr}^*)$.

\subsection{Displacement photocurrent $j_y$ perpendicular to the modulation direction}

The current in the $y$-direction can also be computed in a semiclassical approach.
If a $dc$ electric field is applied in the $y$-direction, the equipotential lines of energy ${\cal E}$
are "meander" lines defined by ${\cal E} = V_n\cos(Qx) -eE_{\rm dc}y$. This gives 
\begin{equation}
   y = {1\over eE_{\rm dc}} [ V_n\cos(Qx)- {\cal E}]
\end{equation}
with an average $y$-value ${\overline y} = - {\cal E}/eE_{\rm dc}$. Quantum-mechanically, we can 
think of these equipotential lines as states. This relation as well as the calculation sketched
in this section is worked out more formally in App. \ref{appendixjy}. Scattering between such meander
states that differ in energy by $\Delta\omega=\omega_c-\omega$ (ignoring the LL energy) involves
jumping a distance $\Delta\omega/eE_{\rm dc}$ in the $y$-direction. Thus, we
can again compute the current by Fermi's golden rule. It is important to observe that the direction 
of the jumps is fixed by the sign of the energy difference. Below, we will comment on the limits
of validity of this approach.

\begin{figure}
 \psfrag{A2}[][][.7]{$\hspace*{.2cm} A_2 $ }
 \psfrag{B2}[][][.7] {$\hspace*{.2cm} B_2 $}
 \psfrag{x}[][][.7] { \hspace*{-.5cm} $\Delta\omega/2V$ }
\epsfig{file=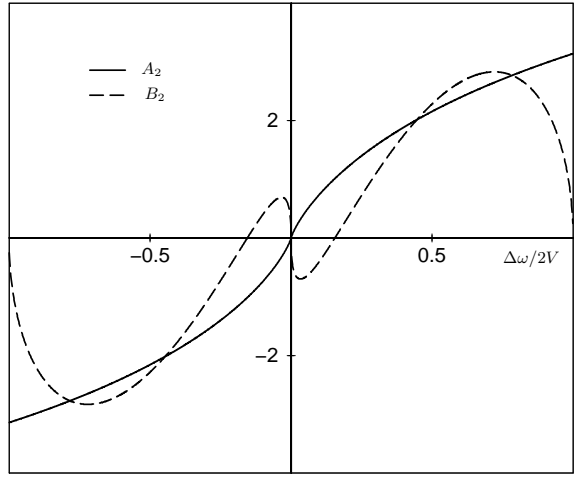,width=7.5cm}
\caption{The functions $A_2(\Delta\omega/2V_N)$ (full line) and $B_2(\Delta\omega/2V_N)$ (dashed line)
describing the dependence of the perpendicular photoconductivity on the microwave frequency, cf.\ 
Eqs.\ (\ref{A2}) and (\ref{B2}).}
\label{fig2}
\end{figure}

The current expression involves the rate of jumps. If $E_{\rm dc}$ is
sufficiently weak, the amplitude of the meander line is very large compared to the scale 
$R_c$ over which jumps occur. On the scale of the jump, the meander lines are   
therefore essentially indistinguishable from the equipotential lines in the absence of the 
$dc$-field. This allows us to employ the rate of jumps which we obtained from the calculation of 
the current in $x$-direction. (We have to set $E_{\rm dc}=0$ in the formulas obtained there, cf.\
Eq.\ (\ref{jumprate})). 
In this way, we obtain for the displacement photocurrent in the $y$ direction
\begin{eqnarray}
\label{photodisplacementy}
   j_y^{\rm photo\,I}
    &=&  {e \over 2\pi \ell_B^2}\left({eER_c\over 4\Delta\omega}\right)^2 
         \sum_n [f_n - f_{n+1}]
         \nonumber\\
         &&\,\,\,\,\,\,\times
        {\Delta\omega\over eE_{\rm dc}}\int {d^2q\over (2\pi)^2} e^{-{q^2\ell_B^2\over 2}}
        [ L_{n + 1} ({q^2\ell_B^2\over 2})
        -L_n({q^2\ell_B^2\over 2})]^2 {\tilde W({\bf q})\over V_n[\sin^2 {Qq_y\over 2} - ({\Delta
        \omega\over
       2V_n})^2]^{1/2}}.
\end{eqnarray}
As mentioned above, we derive this expression more formally in App.\ \ref{appendixjy}.
Computing the integral for white-noise disorder, we obtain the result
\begin{eqnarray}
\label{sigmayyphotoIfinal}
    j_{y}^{\rm photo\,I} &=&  {8\Delta\omega \over (2\pi)^3  V_N  \ell_B^4 
   E_{\rm dc}}\left( {eER_c\over 4\Delta\omega}
    \right)^2  {1\over 2\pi \nu \tau} K(\sqrt{1- (\Delta \omega /2 V_N)^2}),
\end{eqnarray}
valid for $ \lambda_F \ll a \ll R_c$ and $ T\gg V_N  $. Note that the current is {\it not} linear in the 
applied $dc$ electric field but rather diverges as as $1/E_{\rm dc}$. This anomalous behavior
is associated with the fact that the length of the jumps diverges with decreasing $dc$ electric 
field and that the direction of all jumps is the same, fixed by the sign of $\Delta\omega$.
Rewriting the result in terms of a conductivity, we obtain
\begin{equation}
\label{sigmaphotoIyyfinal}
  \sigma_{yy}^{\rm photoI} = \left[ e^2D_{yy}\nu^*\right]\left(aB/\pi\sqrt{\tau_s^*\tau^*_{\rm tr}}
  \over E_{\rm dc} \right)^2
  \left(eER_c\over 4\Delta \omega\right)^2
  A_2(\Delta\omega/2V_N).
\end{equation}
Note that the first factor is just the dark conductivity $\sigma_{yy}$.
The derivation of the result for smooth disorder is sketched in App.\ \ref{smooth}.
The function $A_2(x)$ is defined by
\begin{equation}
\label{A2}
   A_2(x) = 2x K(\sqrt{1- x^2})
\end{equation}
and plotted in Fig.\ \ref{fig2}.
The sign of the photocurrent is given by ${\rm sgn}(\omega_c-\omega)$, as in the case of $\sigma_{xx}^
{\rm photoI}$. 

The magnitude of $\sigma^{\rm photo}_{yy}$ can be understood as follows. Since all jumps are in the same direction, 
we estimate the current density directly. Effectively one LL contributes so that the relevant 
density of electrons is $1/2\pi\ell_B^2$. The step length is $\Delta\omega/eE_{\rm dc}$ and the rate of
jumps is given by $(1/\tau_s^*)(eER_c/\Delta\omega)^2(\tau_s^*/\tau_{\rm tr}^*)$ where the first
two factors are the bare rate and the last factor again reflects the partially destructive interference 
between the two contributions to disorder-assisted microwave absorption. Thus, we find a current 
of order $j_y^{\rm photoI} \sim (e/2\pi\ell_B^2)(\Delta\omega/eE_{\rm dc})
(1/\tau_s^*)(eER_c/\Delta\omega)^2(\tau_s^*/\tau_{\rm tr}^*)$,
in agreement with Eq.\ (\ref{sigmaphotoIyyfinal}).

The limits of validity of this result are most naturally discussed in terms of a 
semiclassical picture. Semiclassically, the individual scattering events such as disorder-assisted
microwave absorption leave the $y$-coordinate of the electron essentially unchanged
(to an accuracy of $R_c$). The full jump by $\Delta\omega/eE_{\rm dc}$ is realized only
if the electron remains in the meander state it scattered into for sufficiently long times to 
explore its entire $y$-range. Under the condition $\tau_s^* \ll \tau_{\rm in}$,
the electron will diffuse on the meander line before equilibrating by inelastic processes.
The typical diffusion distance $\sqrt{D_{yy}\tau_{\rm in}}$ in the $y$-direction should be 
larger than the amplitude $V_N/eE_{\rm dc}$ of the meander line. 
Thus, we find that the condition for the validity of the expression (\ref{sigmaphotoIyyfinal}) is 
\begin{equation}
    E_{\rm dc} \gg E_{\rm dc}^*={aB\over 2\pi\sqrt{\tau^*\tau_{\rm in}}}
\end{equation}
Note that this is just the opposite of the range of validity of the dark conductivity $\sigma_{yy}$
computed in Eq.\ (\ref{sigmadarkyy}). 

For smaller $dc$ electric fields, the jumps are no longer all in the same direction and we can estimate 
the displacement photoconductivity as follows. The disorder-induced microwave absorption excites the electrons 
to a meander (equipotential) line which is shifted in the $y$-direction by $\Delta\omega/eE_{\rm dc}$ relative
to the initial state. For definiteness, assume that this shift is in the positive $y$-direction.
Since quasiclassically, the jump itself leaves the $y$-coordinate of the electrons unchanged (to an accuracy
of $R_c$), the electrons will initially populate only those parts of the excited meander line, which are 
at least a distance $\Delta\omega/eE_{\rm dc}$ from its top (in the $y$-direction).    
After the excitation, the electrons begin to diffuse on the equipotential line due to disorder scattering,
typically a distance $\sqrt{D_{yy}\tau_{\rm in}}$ before they relax back. 
Thus, after time $\tau_{\rm in}$, the population of the excited meander line will extend further in the positive $y$-direction by a distance 
$\sqrt{D_{yy}\tau_{\rm in}}$, and the average {\it positive} drift per electron is 
$[\sqrt{D_{yy}\tau_{\rm in}}/(V_N/eE_{\rm dc})] \sqrt{D_{yy}\tau_{\rm in}}$.
These arguments lead to a displacement photoconductivity of
\begin{equation}
\label{displacementyy-linear}
  \sigma_{yy}^{\rm photoI} \sim [e^2D_{yy}\nu^*] \left(eER_c\over\Delta\omega\right)^2
   \left(\tau_{\rm in}\over \tau_{\rm tr}^*\right)
\end{equation}
valid for $E_{\rm dc} \ll E_{\rm dc}^*$. Note that this result matches with Eq.\ (\ref{sigmaphotoIyyfinal})
for $E_{\rm dc} = E_{\rm dc}^*$. It is interesting to note that even the linear-response
displacement photoconductivity involves the inelastic relaxation time $\tau_{\rm in}$.

\section{The effect of a non-equilibrium electron distribution on the photocurrent}
\label{photoII}

\subsection{Distribution function}

For non-zero inelastic relaxation time $\tau_{\rm in}$, the microwave irradiation changes the
electron distribution function, away from the equilibrium Fermi-Dirac distribution. In this section,
we consider the contribution to the photocurrent arising from this change in $f_{nk}$.

The microwave-induced change in the distribution function can be computed from the stationary 
kinetic equation in the absence of the $dc$ electric field. Elastic disorder scattering contributes
only when states of the same energy have different occupations. Since this is not the case for
$E_{\rm dc}=0$, we can ignore elastic disorder scattering when computing the microwave-induced
change in $f_{nk}$. Thus, the kinetic equation reduces to a balance between the microwave-induced
collision integral and inelastic relaxation which yields to linear order in the microwave intensity
\begin{equation}
    \delta f_{nk} = f_{nk} - f^0_{nk} =
        \tau_{\rm in} \sum_{n'k'}\sum_{\sigma=\pm} 2\pi |\langle n'k' | T_\sigma | nk \rangle |^2
        [ f^0_{n'k'} - f^0_{nk} ] \delta ( \epsilon^0_{nk} - \epsilon^0_{n'k'} + \sigma\omega ).
\end{equation} 
As before, we restrict attention to temperatures $T\gg V$ so that $f^0_{nk}\simeq f_n^0$, independent 
of $k$. Inserting the explicit expression (\ref{photoImatrixelement}) for the $t$-matrix element for white-noise
disorder and 
using the decoupling of the $q$-integration for $\lambda_F\ll a \ll R_c$, we obtain for the change in the 
distribution function
\begin{eqnarray}
\label{deltaf}
        \delta f_{nk}= 2
        \left({eER_c\over 4\Delta\omega}\right)^2 
        \sum_{\sigma=\pm}[ f^0_{n+\sigma} - f^0_{n}]
        {\tau_{\rm in}\over\tau_{\rm tr}^*(\epsilon^0_{nk}-\sigma\Delta\omega)}
        \,\theta(V-|\epsilon^0_{nk}-\sigma\Delta\omega|).
\end{eqnarray} 
Note that $k$ enters this expression only through $\epsilon^0_{nk}$. Strictly speaking, this expression 
breaks down 
when $\epsilon^0_{nk}-\sigma\Delta\omega$ approaches the band edge. In this limit, it is no longer 
sufficient to treat the microwave field to linear order in the intensity. Effectively, this divergence
is cut off when $\delta f_{nk}$ becomes of order unity, i.e., for distances $\Delta\epsilon$ from the 
band edge satisfying
\begin{equation}
\label{cutoff}
   \Delta\epsilon \ll 
   V\left[\left(eER_c\over\Delta\omega\right)^2\left(\tau_{\rm in}\over\tau^*_{\rm tr}\right)\right]^2
\end{equation}
Here, we used that the DOS has a square-root divergence at the band edge. We also note that our linear
approximation in the microwave intensity breaks down completely beyond microwave intensities given by
$(eER_c/\Delta\omega)^2(\tau_{\rm in}/\tau^*_{\rm tr})\sim 1$.

In the estimate
\begin{equation}
\delta f_{nk} \sim
(1/\tau^*_s) (eER_c/\Delta\omega)^2 
(\tau_s^*/\tau_{\rm tr}^*)\tau_{\rm in}
\end{equation}
for the magnitude of $\delta f_{nk}$,
the first three factors are the rate for disorder-assisted microwave absorption and emission. 
The last factor $\tau_{\rm in}$ represents the time interval during which electrons are excited.  
The expression (\ref{deltaf}) will form the basis of our calculation of the distribution-function mechanism for the 
photocurrent to which we now turn. 

\subsection{Distribution-function contribution to the photoconductivity along the modulation direction}

Going through the same steps as in the derivation of the dark current in the $x$-direction, 
one finds that Eq.\ (\ref{sigmadarkxxprel}) for $\sigma_{xx}$ remains valid even for a 
non-equilibrium distribution function, as long as it depends on $k$ through $\epsilon^0_{nk}$ 
only. Thus, we find for the distribution-function contribution to the photoconductivity 
\begin{equation}
    \sigma_{xx}^{\rm photo\,II} =  {e^2N\over L_xL_y}
2\pi\ell_B^2{1 \over 2\pi \nu \tau} \sum_{nk}
   \left(-{\partial \delta f(\epsilon^0
_{nk})\over \partial \epsilon^0_{nk}}\right)
    \nu^*(\epsilon^0_{k}).
\end{equation}
Inserting $\delta f_{nk}$ from Eq.\ (\ref{deltaf}), performing the sum over the LL index $n$, and rewriting
the sum over $k$ as an energy integral involving the density of states, we obtain our result
\begin{eqnarray}
\label{sigmaxxphotoIIfinal}
\sigma^{\rm photo\,II}_{xx}= 4  \left(\tau_{\rm in} \over 
\tau_{\rm tr}^* \right)\left({eER_c\over 4\Delta\omega}\right)^2 
\left[e^2D_{xx}\nu^*\right] B_1(\Delta\omega/ 2V_N)
\end{eqnarray}
for the distribution-function contribution to the photoconductivity. The function
\begin{equation}
\label{B1}
B_1(\Delta\omega/2V_N)=-\frac{\partial} {\partial \Delta\omega} 
\int_{-V_N}^{V_N-|\Delta\omega|} d\epsilon \,\tilde\nu^*(\epsilon+|\Delta\omega|) [\tilde\nu^*(\epsilon)]^2.
\end{equation} 
is of order unity for $\Delta\omega\sim V_N$.
This result shows that for the parallel photocurrent, the distribution-function contribution is larger 
than the displacement contribution (\ref{sigmaxxphotoIfinal})
by a large parameter $\tau_{\rm in}/\tau_s^*$. 
This result is consistent with earlier results for disorder-broadened Landau levels.\cite{Vavilov,Dmitriev2}

The integral (\ref{B1}) has a logarithmic singularity at the lower limit, which has the same origin as  
the divergence as the dark conductivity $\sigma_{xx}$ in Eq.\ (\ref{sigmaepsilondarkxx}). 
Thus, the singularity is cut off in the same way as for the dark conductivity. (For small 
$\Delta\omega$, one may seemingly have a more serious singularity. However, we should remember 
that the DOS arising from $\delta f$ never really becomes singular, cf.\ the discussion above
around Eq.\ (\ref{cutoff}).)
Since the logarithmic singularity dominates the integral (\ref{B1}), we can replace $\epsilon$ by the lower
limit in the DOS $\tilde\nu^*(\epsilon+|\Delta\omega|)$. In this way, we obtain the result
(cf.\ Figs.\ \ref{fig1})
\begin{equation}
\label{B11}
   B_1(x) = {1\over 16}{1-2|x|\over (|x|-|x|^2)^{3/2}}\ln {V_N\over \Delta} {\rm sgn} x.
\end{equation}
Here, $\Delta$ denotes an effective broadening in energy of the band edge, either due to 
disorder or a finite $dc$ electric field, which cuts off the 
logarithmic  singularity. 
Interestingly, this shows that the sign of the photocurrent is
given by ${\rm sgn}(\omega_c-\omega)$ for $|\Delta\omega|<V_N$ only. For $|\Delta\omega|=V_N$, we find additional
sign changes which are associated with the singular nature of the DOS at the Landau level edge, cf.\
Eq.\ (\ref{DOS}).

\subsection{Distribution function contribution to the photocurrent perpendicular to the modulation direction }

To compute the distribution-function contribution to the photoconductivity $\sigma_{yy}$, we note that
Eq.\ (\ref{sigmadarkyyprel}) remains valid for non-equilibrium distribution functions which depend on $k$
through $\epsilon_{nk}^0$, only. Thus, our starting point is
\begin{eqnarray} 
\sigma_{yy}^{\rm photo\,II} = - {e^2\over 2\pi} (2\pi \nu \tau) {1\over L_x L_y}   
\sum_{n k} \left({\partial \epsilon_{nk}^0\over \partial k}\right)^2  
 {1\over \nu^*(\epsilon^0_{nk})}    
 {\partial \delta f_{nk}\over\partial \epsilon_{nk}}.
\end{eqnarray}
Inserting $\delta f_{nk}$ from Eq.\ (\ref{deltaf}) and rewriting the sum over $nk$ as an integral, we 
obtain the final result
\begin{eqnarray}
\label{sigmaphotoIIfinal}
    \sigma_{yy}^{\rm photoII}= 4 \left(\tau_{\rm in}\over\tau_{\rm tr}^*\right)
    \left({eER_c\over  4\Delta\omega}\right)^2[e^2 D_{yy} \nu^*] B_2(\Delta\omega/2V_N).
\end{eqnarray}
The expression (\ref{sigmaphotoIIfinal}) is given in terms of the function
\begin{eqnarray}
\label{B2}
    B_2(\Delta\omega/2V_N)=  - {\partial\over\partial\Delta\omega}
    \int_{-V_N+|\Delta\omega|}^{V_N} d\epsilon {1\over [\tilde\nu^*(\epsilon)]^2}   
    \tilde\nu^*(\epsilon-|\Delta\omega|).
\end{eqnarray}
This integral is elementary and we obtain
(cf.\ Fig.\ \ref{fig2})
\begin{equation}
B_2\left(x\right)
=\left[8 \left|x\right|\left(
\arcsin\left(1- 2\left|x\right| \right)+
\frac{\pi}{2}\right) 
- 8 \;   \sqrt{ \left|x\right|- \left|x\right|^2}\right] {\rm sgn}x . 
\end{equation} 
Asymptotically, this function behaves as  
\begin{equation} 
  B_2(x)  \simeq \left\{ 
    \begin{array}{ccc} - 8 \sqrt{ |x|} {\rm sgn}x & & 
|x| \ll 1 \;,  \\
4\sqrt{1-|x|}{\rm sgn}x & &
1- |x| \ll 1
    \,. \end{array} \right. 
\end{equation}  
Interestingly, $B_2(x)$ has different signs in these two limits, implying that
the function $ B_2(x) $ must have a node between 
the arguments $x=0$ and $x=1$.   
This node is approximately at
$ |\Delta \omega|\approx 2V_N /\pi^2  \approx 0.2 V$. 
As a result, the distribution-function contribution to the photoconductivity 
$ \sigma^{\rm photo\,II}_{yy} $ has the same sign 
as $ \sigma^{\rm photo\,I}_{yy} $ only in the range 
$ |\Delta \omega| \gtrsim  0.2 V_N$. 
Again, we associate this behavior with the anomalous behavior of the DOS.
 
The magnitude is, apart from a function of $\Delta\omega/V_N$, of order $\sigma_{yy}\delta f$. 
Remarkably, in this case the magnitude is of the same order as the displacement 
contribution in Eq.\ (\ref{displacementyy-linear}). The reason for this is that 
the displacement mechanism exhibits a singular, non-Ohmic dependence on the $dc$ electric field 
which is cut off at small fields by inelastic processes only. 
A more detailed analysis beyond this order
of magnitude comparison would require an accurate calculation of the displacement contribution
to the photocurrent in the linear-response regime. Such a calculation is beyond the scope of this 
paper and the result may in any case be sensitive to details of the model for inelastic relaxation. 

\subsection{Weiss oscillations of the photocurrent}
\label{Weiss}

The Weiss oscillations arising from the oscillatory behavior of the $V_n$ as function of
LL index $n$ or magnetic field
have two effects on the photocurrent. First, they lead to a modulation of the 
amplitude of the photocurrent. This amplitude modulation is similar to that of the dark conductivity 
as the photoconductivity is proportional to the dark conductivity. A difference may
arise from the additional prefactor $\tau_{\rm in}/\tau_{\rm tr}^*$ entering the photocurrents,
cf.\ Eqs.\ (\ref{sigmaxxphotoIIfinal}) and (\ref{sigmaphotoIIfinal}). Specifically, if the ineleastic 
relaxation rate $\tau_{\rm in}$
depends in a different way on the LL DOS $\nu^*$ compared to the transport time $\tau_{\rm tr}$, there may
be a distinct difference between the Weiss oscillations in the dark and the photoconductivity. 
Depending on whether the Weiss oscillations in the photoconductivity are more or less pronounced than 
those in the dark conductivity, this may help or impede reaching negative conductivities and hence 
observing the zero-resistance state. 

A second effect of the Weiss oscillations is associated with the modulation of the LL width $V_n$.
The expressions for the photocurrent (\ref{sigmaxxphotoIIfinal}) and (\ref{sigmaphotoIIfinal})
involve a factor which depends on the ratio of the 
detuning $\Delta\omega$ and the LL width $V_N$ at the Fermi energy. This implies that in the 
limit of well-separated LLs considered here, the range in the detuning over which there is a significant
photocondictivity also oscillates with magnetic field or Fermi energy. 

In the photoconductivity, the $1/B$-periodic Weiss oscillations are superimposed on the microwave-induced 
oscillations which are also periodic in $1/B$. The periods in $1/B$ of these oscillations are 
$ea/mv_F$ and $e/m\omega$, respectively, which can be of comparable magnitude. 

\section{Polarization dependence of the photoconductivity}
\label{apol}

In this section, we discuss the dependence of the 
photoconductivity on the polarization of the microwave field $ {\bf E} $. 
We begin by calculating the transition matrix elements 
for a microwave field polarized in the $y$-direction. 
In this case
\begin{equation}
{\phi} = -{eE\over 2} y 
(e^{i\omega t} + e^{-i\omega t}) = \phi_{+} e^{-i\omega t} + \phi_{-}
e^{i\omega t} \,.  
\end{equation} 
The matrix elements of the operator $ {\phi}_{\pm} $ are given by 
\begin{eqnarray}
\langle n \pm m k'|{\phi}_{\pm}| n k\rangle & = & -
\left(\frac{1}{i} \frac{\partial}{\partial k} \delta(k'-k) \right)
e^{-(k-k')^2\ell_B^2/4}  \nonumber \\
& & \left\{ 
\delta_{m,0} L_n^0\left(\frac{(k'-k)^2 \ell_B^2}{2}\right) 
\mp  \delta_{m,\pm 1} \frac
{(k'-k) \ell_B}{\sqrt{2n}} 
L_n^1\left(\frac{(k'-k)^2 \ell_B^2}{2}\right)\right\}  
\end{eqnarray}
for $ m \ge 0 $.
Using $ L^1_n(0)=1 $ and $ L^1_n(0)=n+1 $, 
we obtain for the transition matrix element ${T}_\pm $ 
the large-$n$ result 
\begin{equation}
   \langle n\pm 1 k' | {T}_\pm | n k \rangle =  
  -  \frac{1}{i}   \left({eER_c\over 4 \Delta \omega}\right) 
[ \langle n \pm 1 k' |U|n \pm 1 k\rangle 
-\langle n k'|U|n k\rangle ] 
\end{equation}
valid for $ \Delta \omega\ll \omega_c $.
This matrix element differs from the corresponding 
matrix element for polarization in the $x$-direction by a  
multiplicative prefactor of unit modulus. As the 
photoconductivity depends only on the modulus of the transition matrix 
elements, we find that the photoconductivity is the same for microwave fields
linearly polarized in the $x$ and $y$-directions. One also readily concludes that
more generally, the photoconductivity remains unchanged for any linear polarization 
of the microwaves. 

We now turn to irradiation by circularly polarized 
microwave fields which are described by 
\begin{equation}
 \phi_{\sigma_{\pm}} = -{eE\over 2 \sqrt{2}}
 \left((x\pm iy)e^{-i\omega t}+(x \mp iy)e^{i\omega t}\right) \,.
\end{equation}
Combining the transition matrix elements for microwaves linearly polarized 
in the $x$ and $y$-directions, one finds zero photoconductivity for $ \phi_{\sigma_{+}}$. 
In this case, the $ {\bf E} $-vector rotates opposite  
to the circular cyclotron motion of the electrons in the magnetic field.   
For $\phi_{\sigma_{-}}$, both ${\bf E}$ and the cyclotron motion rotate in the same direction and  
the photoconductivity is double that for
linearly polarized microwave fields. 

\section{Conclusions}
\label{conclusions}

We have computed the microwave-induced photocurrent in the regime of high Landau level filling 
factors, in a model in which the Landau levels are broadened into a band due to a static periodic modulation. 
Assuming that the static modulation is small compared to the spacing between LLs, the eigenstates 
are still given by the Landau level oscillator states in the Landau gauge. The localization
properties of these states allow us to compute the dark conductivity as well as the microwave-induced 
photoconductivity using Fermi's golden rule. The Fermi's golden rule expression for the current 
directly suggests that there are two distinct mechanisms contributing to the photocurrent, analogous to 
previous results \cite{Vavilov,Dmitriev2}
for disorder-broadenend Landau levels. (i) The 
displacement 
mechanism relies on the spatial displacements associated with disorder-assisted microwave absorption and 
emission. This contribution can be associated with an additional, microwave-induced contribution to the 
transition matrix element in the Fermi's golden rule expression. (ii) The distribution-function mechanism 
by contrast, relies on the microwave-induced change in the electronic distribution function, again due to 
disorder-assisted microwave absorption and emission. 

For the photocurrent parallel to the modulation direction, we find that the 
distribution-function mechanism [cf.\ Eq.\ (\ref{sigmaxxphotoIIfinal})] dominates by a large factor $\tau_{\rm in}/\tau_s^*$
over the displacement contribution [cf.\ Eq.\ (\ref{sigmaxxphotoIfinal})], 
in agreement with earlier 
results for disorder broadened Landau levels.\cite{Vavilov,Dmitriev2} 
The sign of the photocurrent changes with the sign of the 
detuning $\Delta\omega=\omega_c-\omega$ of the microwaves. For the dominant distribution function mechanism,
there are additional sign changes associated with the divergence of the density of states at the 
edge of the Landau level. Remarkably, the situation is rather different for the transverse photocurrent 
perpendicular to the 
modulation direction. In this case, we find that the displacement mechanism is in some sense
singular with the result that both contributions to the photocurrent [cf.\ Eqs.\ (\ref{sigmaphotoIyyfinal}) 
and (\ref{sigmaphotoIIfinal})] are of the same order. 
We find that our results remain unchanged for any linear microwave polarization. For circular polarization, 
we find a nonzero photoconductivity only when the microwave electric field rotates in the same direction as
the cyclotron rotation of the electrons in the magnetic field.

\acknowledgments

We thank M.E. Raikh for illuminating discussions. 
LIG thanks the Free University of Berlin and 
FvO the Weizmann Institut for hospitality (supported by LSF and the Einstein Center)
while part of this work has been performed. 
This work has been supported by NSF grants DMR02-37296 and EIA02-10736 (LIG), the Institut Universitaire
de France (FWJH), 
the DFG-Schwerpunkt Quanten-Hall-Systeme (JD and FvO), and the Junge Akademie (FvO).

\appendix

\section{Smooth disorder potentials}
\label{smooth}

As opposed to a white-noise potential, the (zero-magnetic-field) single particle time 
$\tau_s$ is different from the transport time $\tau_{\rm tr}$ for a smooth disorder potential. Specifically, 
the single-particle time can be expressed in terms of the correlator $W$ as 
\begin{equation}
   {1\over \tau_s} = 2\pi \sum_{\bf q}
      \tilde W({\bf q}) \langle \delta (\epsilon_k - \epsilon_{k+q})\rangle_{\rm FS}
     = {1\over \pi v_F} \int_0^\infty dq \tilde W({\bf q}),
\end{equation}
where the average is over the Fermi surface and $\epsilon_k$ denotes the zero-field dispersion.
Likewise, the transport time can be expressed as 
\begin{equation}
   {1\over \tau_{\rm tr}} = 2\pi \sum_{\bf q}(1-\cos \theta_{\bf q})\tilde W({\bf q}) \langle \delta (\epsilon_k - \epsilon_{k+q})\rangle_{\rm FS}
     = {1\over \pi v_F} \int_0^\infty dq (q^2/2k^2_F)\tilde W({\bf q}),    
\end{equation}
where $\theta_{\bf q}$ denotes the scattering angle. Note that $\tau_{\rm tr}/\tau_s\sim (k_F\xi)^2$.

We start by considering the elastic scattering times for smooth disorder. For $\tau_s^*$, we need to reconsider the ${\bf q}$-integration 
in Eq.\ (\ref{Laguerreintegral}),
\begin{equation}
   I_0=\int {d^2q\over (2\pi)^2}e^{-q^2\ell_B^2\over 2} [L_n({q^2\ell_B^2\over 2})]^2 \tilde W({\bf q})
    \delta(\epsilon^0_{nk}-\epsilon^0_{nk+q_y}).
\end{equation}
In the limit of large $N$, the Laguerre polynomial has oscillations on the $q$-scale of $\lambda_F/\ell_B^2$ 
and falls off on the scale
of $R_c/\ell_B^2$. The correlator falls off on the scale $1/\xi$ and finally, the characteristic scale of the 
argument of the $\delta$-function is $a/\ell_B^2$. Thus, unlike for white-noise potential, it is now the correlator 
$\tilde W({\bf q})$ which cuts off the integral at large $q$. Under the condition 
$\lambda_F \ll a \ll \ell_B^2/\xi$, we can still factorize the $q$-integration as 
\begin{equation}
   I_0=\int {d^2q\over (2\pi)^2} e^{-q^2\ell_B^2\over 2} [L_n({q^2\ell_B^2\over 2})]^2 \tilde W({\bf q})
    \langle \delta(\epsilon^0_{nk}-\epsilon^0_{nk'})\rangle_{k'}.
\end{equation}
Since the integral is now cut off at large $q$ by $W$, it is sufficient to employ the semiclassical equation
\begin{equation} 
    e^{-q^2\ell_B^2\over 4} L_n({q^2\ell_B^2\over 2}) \simeq \sqrt{2\over\pi qR_c}
       \cos(qR_c-\pi/4)
\end{equation}
in the remaining integral.
(Strictly speaking, we need a slightly more accurate approximation. However, this changes only the argument of
the cosine \cite{Erdelyi} which does not affect the results.)
This yields 
\begin{equation}
   I_0={1\over 2\pi^2 R_c}\int_0^\infty dq  \tilde W({\bf q})\langle                                                          \delta(\epsilon^0_{nk}-\epsilon^0_{nk'})\rangle_{k'} 
    ={\nu^*(\epsilon)\over \nu}{1\over 2\pi\tau_s}.
\end{equation}
The same integral is involved in the computation of the dark conductivity $\sigma_{yy}$.

The transport time involves the integral 
\begin{equation}
   I_2=\int {d^2q\over (2\pi)^2} {q^2\over 2k_F^2}e^{-q^2\ell_B^2\over 2} [L_n({q^2\ell_B^2\over 2})]^2 \tilde W({\bf q})
    \delta(\epsilon^0_{nk}-\epsilon^0_{nk+q_y}),
\end{equation}
where we used that $1-\cos\theta_{\bf q}\simeq q^2/2k_F^2$ for $q\ll k_F$. An analogous analysis as for $I_0$ above
gives the result
\begin{equation}
   I_2={\nu^*(\epsilon)\over \nu}{1\over 2\pi\tau_{\rm tr}}.
\end{equation}
The same integral appears in the calculation for the dark conductivity $\sigma_{xx}$.

A similar integral appears in the calculation for the displacement photocurrent $\sigma_{yy}^{\rm photoI}$, namely
\begin{equation}
   J_0=\int {d^2q\over (2\pi)^2} e^{-q^2\ell_B^2\over 2} [L_{n+1}({q^2\ell_B^2\over 2})
     -L_{n}({q^2\ell_B^2\over 2})]^2 \tilde W({\bf q})
    \delta(\epsilon^0_{nk}-\epsilon^0_{nk+q_y}).
\end{equation}
For smooth disorder, the difference of Laguerre-polynomials is suppressed relative to a single 
Laguerre-polynomial. Using that $qR_c^{(n+1)}\simeq qR_c^{(n)}+q/k_F$, we find that the difference 
effectively introduces a small factor $q^2/k_F^2$ into the integrand and thus (cp.\ the calculation 
for $I_2$ above)
\begin{equation}
    J_0={\nu^*(\epsilon)\over \nu}{1\over \pi\tau_{\rm tr}}.
\end{equation}
For the displacement photocurrent $\sigma_{xx}^{\rm photoI}$, we need to consider
\begin{equation}
   J_2=\int {d^2q\over (2\pi)^2} (q^2/2k^2_F)  e^{-q^2\ell_B^2\over 2} [L_{n+1}({q^2\ell_B^2\over 2})
     -L_{n}({q^2\ell_B^2\over 2})]^2 \tilde W({\bf q})
    \delta(\epsilon^0_{nk}-\epsilon^0_{nk+q_y}).
\end{equation}
Again, the difference of Laguerre polynomials introduces a small factor $q^2/k_F^2$ into the integrand. 
The resulting integral can no longer be related directly to either $\tau_{\rm tr}$ or $\tau_s$. However,
noting that every factor $(q/k_F)^2$ reduces the integral by a factor of order $1/(k_F\xi)^2$, we can 
estimate\begin{equation}
    J_2 \sim {\nu^*(\epsilon)\over \nu}{1\over \pi\tau_{\rm tr}}{1\over (k_F\xi)^2}
       \sim {1\over \pi\tau^*_{\rm tr}}{\tau_s^*\over \tau^*_{\rm tr}}.
\end{equation}
The precise numerical prefactor depends on the detailed nature of the smooth potential. 

\section{Distribution function for large $dc$ electric fields}
\label{heating}

For large $dc$ electric fields, $E_{\rm dc}\gg E_{\rm dc}^*$, Joule heating effects become important. 
In this appendix, we study the distribution function in this limit. We start by decomposing the 
distribution function $f_{nk}$ into symmetric and antisymmetric parts $\sigma_{nk}$ and $\alpha_{nk}$
under $k\to -k$, i.e.\
\begin{equation}
   f_{nk}=\sigma_{nk}+\alpha_{nk}.
\end{equation}
(Recall that the electron dispersion is symmetric under this transformation.)
Specifically, we can write $\sigma_{nk} = (1/2)[f_{nk}+f_{n-k}]$ and $\alpha_{nk}=(1/2)[f_{nk}-f_{n-k}]$.
Inserting this decomposition into the kinetic equation in the presence of a $dc$ electric field in the 
$y$-direction (and without microwaves), we obtain the two equations
\begin{eqnarray}
   -eE_{\rm dc} {\partial \alpha_{nk}\over\partial k} &=& {\sigma_{nk}-\sigma_{nk}^0\over \tau_{\rm in}}
   \nonumber\\
   -eE_{\rm dc} {\partial \sigma_{nk}\over\partial k} &=& {\alpha_{nk}\over \tau_s^*(\epsilon_{nk})}.   
\end{eqnarray}
Here, we have used the inequality $\tau_{\rm in} \gg \tau_s^*$ and the fact that the disorder collision integral 
vanishes for the symmetric part $\sigma_{nk}$. In addition, we have rewritten the collision integral for the 
antisymmetric part as $(\partial \alpha/\partial t)_{\rm dis}=-\alpha_{nk}/\tau_s^*(\epsilon_{nk})$. Inserting the
second of these equations into the first, we obtain
\begin{equation}
\label{sigmaheating}
   (eE_{\rm dc})^2\tau_s^*(\epsilon_{nk}){\partial^2 \sigma_{nk}\over \partial k^2} = 
      {\sigma_{nk}-\sigma_{nk}^0\over \tau_{\rm in}}.
\end{equation}
Note that this equation reproduces the estimate of the characteristic electric field $E_{\rm dc}^*$. This 
follows immediately from the fact that the characteristic scale of the $k$ dependence is $a/\ell_B^2$.

For $E_{\rm dc}\ll E_{\rm dc}^*$, we therefore find $\sigma_{nk}\simeq\sigma_{nk}^0$. This is the starting 
point of the calculation leading to the expression (\ref{sigmadarkyy}) for the dark conductivity $\sigma_{yy}$. 
In the opposite limit $E_{\rm dc}\gg E_{\rm dc}^*$, we write $\sigma_{nk}={\overline\sigma}_n+\delta\sigma_{nk}$, 
where ${\overline\sigma}_{n}$ is the average of $\sigma_{nk}$ over $k$. Note that $\alpha_{nk}$ which determines 
the current and hence the conductivity is directly related to $\delta\sigma_{nk}$. Then, Eq.\ 
(\ref{sigmaheating}) shows that in magnitude $\delta\sigma_{nk} \sim (E^*_{\rm dc}/E_{\rm dc})^2\sigma_{nk}^0$.
As a result, we expect that heating reduces the dark conductivity according to Eq.\ (\ref{heatingsigmayy})
in Sec.\ \ref{dark}.

\section{Explicit calculation of displacement photocurrent $j_y$}
\label{appendixjy}

In this appendix, we derive the displacement photocurrent in the $y$-direction more formally. 
In order to derive the quantum version of the meandering equipotential lines, we consider the 
Schr\"odinger equation, including the $dc$ electric field in the $y$-direction, in LL representation
\begin{eqnarray}
   \langle nk | H_0 | n'k'\rangle = \epsilon^0_{nk} \delta_{nn'}\delta_{kk'}
      -eE_{\rm dc} (-i)\left( {\partial\over \partial k}\delta_{kk'}\right)
       f_{nn'} (k-k')
\end{eqnarray}
with
\begin{equation}
   f_{nn'}(k-k') = \left\{ \begin{array}{ccc} \left(2^n n!\over 2^{n'}n'!\right)^{1/2}
       (k-k')^{n'-n} e^{-(k-k')^2/4} L_n^{n'-n}\left({(k-k')^2\over 2}\right)  & {\rm if} & n'\geq n \\
       \left(2^{n'} n'!\over 2^{n}n!\right)^{1/2}
       (k-k')^{n-n'} e^{-(k-k')^2/4} L_{n'}^{n-n'}\left({(k-k')^2\over 2}\right)  & {\rm if} & n \geq n'
     \end{array} \right. .
\end{equation}
Neglecting LL mixing, we find the Schr\"odinger equation in the quasiclassical limit
\begin{equation}
   \epsilon^0_{nk} \psi_{nk} + eE_{\rm dc} (-i){\partial\over \partial k} \psi_{nk} = {\cal E} \psi_{nk}
\end{equation}
with
\begin{equation}
   |n{\cal E}\rangle = \sum_k \psi_{nk}| nk\rangle .
\end{equation}
This is readily solved and gives the quasiclassical meander states
\begin{equation}
   \psi_{nk} = \psi_{0n} \exp\left\{i\int_0^k dk'{ {\cal E} - \epsilon^0_{nk'} \over eE_{\rm dc}} \right\}
\end{equation}
To count the number of such states, we assume periodic boundary conditions in the $x$-direction, 
$\psi_{nk+L_x/\ell_B^2}=\psi_{nk}$, and thus (with $l\in {\bf Z}$)
\begin{equation}
   {\cal E}_l = {2\pi l eE_{\rm dc} \ell_B^2 \over L_x},
\end{equation}
where ${\cal E}_l$ is measured relative to the LL energy. As the energies ${\cal E}_l$ fall into the range
$eE_{\rm dc}L_y$, the total number of states is ${L_xL_y/ 2\pi\ell_B^2}$, in agreement with the 
LL degeneracy.  Requiring 
$ 1= \langle n{\cal E}_l|n{\cal E}_l\rangle $, we find the normalized meander states
\begin{equation}
   \psi_{nk} = \langle nk|n{\cal E}_l\rangle = \sqrt{2\pi \ell_B^2\over L_xL_y}\exp\left\{i\int_0^k dk'{ {\cal E}
     - \epsilon^0_{nk'} \over eE_{\rm dc}} \right\}.
\end{equation}
To verify that these states are indeed the meander states, we evaluate the expectation value of 
$y=i\partial/\partial k$, and find
\begin{equation}
    \langle n {\cal E}_l | y | n {\cal E}_l\rangle = {2\pi \ell_B^2\over L_xL_y} \sum_k \left( -{{\cal E}_{nl}
      -\epsilon^0_{nk}\over e E_{\rm dc}}\right) = - {{\cal E}_{l} \over eE_{\rm dc}}
\end{equation}
with $ {\cal E}_{nl} = \omega_c (n + 1/2) + {\cal E}_l$, in agreement with the classical expectation.

Following the same arguments as for the displacement photocurrent in the $x$-direction, the current in 
the $y$-direction can now be expressed as
\begin{eqnarray}
    j_y^{\rm photo\, I} 
     = {e\over L_xL_y} \sum_{nn'}\sum_{{\cal E}_l,{\cal E}_{l'}} |\gamma^\omega_{n{\cal E}_l\to n'{\cal E}_{l'}}|^2
     ({\overline y}({\cal E}_{l'}) - {\overline y}({\cal E}_l ))
     [ f({\cal E}_{nl})-f( {\cal E}_{n'l'} )  ] \delta({\cal E}_{nl} -{\cal E}_{n'l'} -\omega),
\end{eqnarray}
where the transition matrix element is given by
\begin{equation}
|\gamma^\omega_{n{\cal E}_l\to n'{\cal E}_{l'}}|^2=
   2\pi  |\langle n'{\cal E}_{l'}| T_\sigma |n {\cal E}_{l}\rangle|^2.
\end{equation}
By carrying out the 
summation over $ n $, $n'$ for $\omega_c\gg T \gg V $ we get
\begin{eqnarray}
   j_y^{\rm photo\,I} 
     = {e\over L_xL_y} {\Delta\omega\over eE_{\rm dc}}\sum_{{\cal E}_l,{\cal E}_{l'}}
    |\gamma^\omega_{N{\cal E}_l\to N+1{\cal E}_{l'}}|^2 \delta({\cal E}_{l} -{\cal E}_{l'} +\Delta\omega).
\end{eqnarray}

The relevant transition matrix element is
\begin{eqnarray}
    |\gamma^\omega_{N{\cal E}_l\to N+1{\cal E}_{l'}}|^2 = 2\pi  \left( {eER_c\over 4\Delta\omega} \right)^2
      | \langle N{\cal E}_l|U|N{\cal E}_{l'}\rangle
      -\langle N+1{\cal E}_l|U|N+1{\cal E}_{l'}\rangle|^2.
\end{eqnarray}
After disorder averaging, the matrix element becomes 
\begin{eqnarray}
      && | \langle N{\cal E}_l|U|N{\cal E}_{l'}\rangle
      -\langle N+1{\cal E}_l|U|N+1{\cal E}_{l'}\rangle|^2
      \nonumber\\
    &&\,\,\,\,\,\,\,\,\,\,\,
    = {1\over 2\pi \nu \tau} \int {d^2q\over (2\pi)^2} |\sum_k
      \psi_{k+q_y}^{({\cal E}_l)*}\psi_k^{({\cal E}_{l'})} e^{iq_x(k+q_y/2)}|^2
      e^{-q^2/2} [L_{N+1}(q^2/2)-L_N(q^2/2) ]^2
\end{eqnarray}
Inserting this into the expression for the current, we obtain
\begin{eqnarray}
    j_y^{\rm photo\,I} &=& {e\over L_xL_y} {2\pi\Delta\omega\over eE_{\rm dc}}\left( {eER_c\over 4\Delta\omega}
    \right)^2  {1\over 2\pi \nu \tau} \int {d^2q\over (2\pi)^2} e^{-q^2/2} [L_{N+1}(q^2/2)-L_N(q^2/2) ]^2
     \nonumber\\
      &&\,\,\,\,\,\,\,\,\,\,\,\,\,\,\,\,
      \times\sum_{{\cal E}_l,{\cal E}_{l'}}|\sum_k
      \psi_{k+q_y}^{({\cal E}_l)*}\psi_k^{({\cal E}_{l'})} e^{iq_x(k+q_y/2)}|^2
      \delta({\cal E}_{l} -{\cal E}_{l'} +\Delta\omega)
\end{eqnarray}
Performing the energy sums yields
\begin{eqnarray}
     \sum_{{\cal E}_l,{\cal E}_{l'}}|\sum_k \psi_{k+q_y}^{({\cal E}_l)*}\psi_k^{({\cal E}_{l'})}
        e^{iq_x(k+q_y/2)}|^2\delta({\cal E}_{l} -{\cal E}_{l'} +\Delta\omega)
     = {L_y\over e E_{\rm dc} } |\int {dk\over 2\pi} e^{{iV\over eE_{\rm dc}}\int_k^{k+q_y}d\tilde k
       \cos(Q\tilde k)+{i\Delta\omega k\over eE_{\rm dc}}+iq_x(k+{q_y\over 2})} |^2.
\end{eqnarray}
The $k$-sum can be turned into an integral which in the limit $E_{\rm dc}\to 0$ can be evaluated in 
the stationary-phase approximation. This gives the result
\begin{eqnarray}
     \sum_{{\cal E}_l,{\cal E}_{l'}}|\sum_k \psi_{k+q_y}^{({\cal E}_l)*}\psi_k^{({\cal E}_{l'})}
        e^{iq_x(k+q_y/2)}|^2\delta({\cal E}_{l} -{\cal E}_{l'} +\Delta\omega)
        ={L_xL_y\over (2\pi)^2 V  \ell_B^2}{1\over [\sin^2 {Q q_y\over 2} -
      ({\Delta\omega\over 2V})^2 ]^{1/2}}
\end{eqnarray}
Note that $E_{\rm dc}$ drops out of this expression. This can be interpreted as follows.
The length over which the electron can jump between meander lines is proportional to
$1/E_{\rm dc}$. On the other hand, the electron density along the meander line is
proportional to $E_{\rm dc}$. Thus, the overall probability to jump is independent of
the $dc$ electric field. In this way, we finally arrive at
\begin{eqnarray}
    j_y^{\rm photo\,I} = {\Delta\omega\over 2\pi V  \ell_B^2 E_{\rm dc}}\left( {eER_c\over 4\Delta\omega}
    \right)^2  {1\over 2\pi \nu \tau} \int {d^2q\over (2\pi)^2} e^{-q^2/2} [L_{N+1}(q^2/2)-L_N(q^2/2) ]^2
     {1\over [\sin^2 {Q q_y\over 2} - ({\Delta\omega\over 2V})^2 ]^{1/2}}
\end{eqnarray}
for the displacement photocurrent in the $y$-direction. Up to the sums over Landau levels, this is just 
the expression for the displacement photocurrent in the $y$-direction in Eq.\ (\ref{photodisplacementy}).

\end{document}